\documentclass[a4paper]{spie}  
\usepackage[]{graphicx}

\usepackage{graphicx}
\usepackage[space]{grffile}
\usepackage{latexsym}
\usepackage{amsfonts,amsmath,amssymb}
\usepackage{url}
\usepackage[utf8]{inputenc}
\usepackage{fancyref}
\usepackage{hyperref}
\hypersetup{colorlinks=false,pdfborder={0 0 0},}

\title{The first SPIE software Hack Day}

\author{S. Kendrew\supit{1}, C. Deen\supit{2}, N. Radziwill\supit{3}, S. Crawford\supit{4}, J. Gilbert\supit{1}, M. Gully-Santiago\supit{5}, P. Kub\'{a}nek\supit{6}
\skiplinehalf
\supit{1}University of Oxford, Department of Physics, Denys Wilkinson Building, Keble Road, Oxford OX1 3RG, United Kingdom\\
\supit{2}Max-Planck-Institut f{\"{u}}r Astronomie, K\"{o}nigstuhl 17, 69117 Heidelberg, Germany\\
\supit{3}James Madison University, Harrisonburg VA 22802, USA\\
\supit{4}South African Astronomical Observatory, Observatory Road, Cape Town, South Africa\\
\supit{5}Department of Astronomy, University of Texas at Austin, 1 University Station C1400, Austin TX 78712, USA\\
\supit{6}Institute of Physics, Czech Republic}

\authorinfo{Send all correspondence to S. Kendrew, sarah.kendrew@astro.ox.ac.uk}

\begin{document} 
  \maketitle 

\begin{abstract}
We report here on the software Hack Day organised at the 2014 SPIE conference on Astronomical Telescopes and Instrumentation in Montr\'{e}al. The first ever Hack Day to take place at an SPIE event, the aim of the day was to bring together developers to collaborate on innovative solutions to problems of their choice. Such events have proliferated in the technology community, providing opportunities to showcase, share and learn skills. In academic environments, these events are often also instrumental in building community beyond the limits of national borders, institutions and projects. We show examples of projects the participants worked on, and provide some lessons learned for future events.

\end{abstract}



\section{INTRODUCTION}
\label{sec:intro}  

The bi-annual SPIE conference on Astronomical Telescopes and Instrumentation is a popular event in the instrumentation and technology development fields of astronomy. A dedicated conference on Software and Cyberinfrastructure is a regular part of the programme. For the first time, the 2014 conference hosted a software Hack Day under this header. Because the community of software engineers and technology managers who attend this conference have regularly discussed and analyzed ways to stimulate and increase collaboration across observatory boundaries, introduction of a Hack Day seemed to be a reasonable next step to bring these potentials into being\cite{chiozzi07, chiozzi08, shortridge08}.  

Hack Days, also known as hackathons or hackfests, are intense immersive events where multi-skilled teams work on collaborative projects. They are often organised around a certain topic or programming language, or focused on a particular subject, problem or challenge (e.g. healthcare\footnote{\url{http://www.nhshackday.com}}, civic activism\footnote{\url{http://hack4good.io}}). They are limited in time, e.g. one day or a weekend, and sometimes have a competitive component, with prizes given to the best hacks. Common technological themes are innovative visualization of large datasets, building applications with APIs to web services, or hardware-based projects. Hackathons have become important networking events in the software and technology industries, as they provide an opportunity for designers, developers or managers to showcase their skills to a wider group they might not otherwise come in contact with\cite{leckart}.

In astronomy, the first Hack Days were organised as part of the .Astronomy conference series. The .Astronomy conferences\cite{dotastro}, which started in Cardiff (UK) in 2008, aim to bring together scientists, educators and outreach professionals, to discuss novel ways of exploiting web-based technologies for astronomy research or public engagement. Since 2009 .Astronomy has included a Hack Day, and in recent years astronomy-themed Hack Days have taken place at the annual conferences of the American Astronomical Society and the UK National Astronomy Meeting.

With an increasing number of robotic telescopes coming online and the advent of mega-facilities such as Gaia, the Large Synoptic Survey Telescope (LSST), and the Square Kilometer Array, control strategies, software and intelligent data processing form an increasingly important part of astronomical observatories. With the SPIE Hack Day, our aim was to provide an opportunity for instrumentation professionals to share skills and collaborate outside of the normal boundaries of an instrumentation project. 

\section{ON THE DAY}
The Hack Day took place on Thursday 26 June over an entire day. Whilst we gathered contact details in the weeks beforehand via an online sign-up form, there was no formal registration procedure, the event was open and free of charge to all conference participants. Lunch and refreshments were provided free of charge.

The format was unscheduled, with no formal presentations, as is traditional for such events. At the start of the day, we gave participants the opportunity to introduce themselves and present what ideas or problems they were interested in working on during the event. Likewise, the day was ended with the participants describing and showcasing the hacks they worked on during the day. In the following section, we highlight some selected Hack Day projects. The names listed alongside the hacks are the participants who proposed and led the project, however many hacks were team-based efforts.

\section{HACK DAY PROJECTS}

\subsection{Sonification of astronomical spectra (J. Gilbert)}

A frequent theme at astronomy hack days is to discover innovative ways of representing astronomical datasets to aid data discovery. One non-traditional avenue is to convert data to sound. This hack aimed to convert astronomical spectra into music; the ability to convert data from a visual into an auditory medium , as well as presenting a creative challenge, can be useful for communicating astronomy to visually impaired people.

The main challenge was to discover how to link the typical characteristics of spectra (e.g. absolute wavelength range, important lines, continuum level vs transients etc.) to the main aspects of a musical piece (time signature, tempo, key, major/minor, pitch); and to do so in a robust way so that spectra which look similar will also \emph{sound} similar.  James wrote a number of routines to quantify these characteristics and link them directly to the musical parameters; then generated a short, loopable tune.  An important goal is to make sure we can definitely `hear' the emission/absorption lines of the object, and for the sound to be attractive.

\subsection{A telescope control system for Android platforms (P. Kub\'{a}nek)}

RTS2, or Remote Telescope System, 2nd Version, is an open source observatory control software. It allows astronomers around the world to let the machines control their observatories. So they can sleep during the night and work during the day. RTS2 takes care of all duties of a (night) observer - it opens and closes observatory protected cover at
evening, morning and in case of bad weather, command telescope to point, filter wheel to turn and camera to expose. It select the observations from the database, and logs the observations to the database. It is used on more than 20 observatories around the planet, making it one of the most popular observatory control programmes.

During the Hack Day, RTS2 creator Petr Kub\'{a}nek worked on developing an Android application for the software.

\subsection{Making figures from a textbook interactive with MPLD3 and d3.js (M. Gully-Santiago)}

A growing theme in modern astrophysics is the growth of the Volume, Variety, and Velocity of astronomical data \cite{garrett14}. Particularly, the increasing multi-dimensionality of large astronomical datasets is giving rise to new data visualization techniques\cite{glue}.  Michael Gully-Santiago's hack was aimed at transforming static figures to interactive figures for web browsers.  The starting static figure was a single panel of Figure 10.21 of the textbook ``Statistics, Data Mining, and Machine Learning in Astronomy: A Practical Python Guide'' \cite[hereafter ICVG]{icvg}.  The enhanced, interactive figure had similar axes and data, but with interactive tooltips so that viewers could hover over individual data points to see all available information about individual sources.  

The figure was made interactive by modifying the original source code with the Python module MPLD3\footnote{ \url{http://mpld3.github.io/}}, which converts matplotlib graphics directives into SVG javascript commands in d3.js\footnote{Data Driven Documents \url{http://d3js.org/}}.  The outcome is hosted on the SPIE hack day GitHub repository webpage\footnote{\url{http://spie-hack-day-2014.github.io/astroML/}}.  

The hack has catalyzed a continued effort to enhance a selection of book figures from ICVG2014, which is now hosted at \url{gully.github.io/astroMLfigs}.  The enhancements go beyond adding d3.js interactivity, and include IPython notebooks with step-by-step guides to how the authors generated the complex figures in the textbook.  The enhanced figures could become a community resource for textbook readers.

\subsection{OKGalileo: An online dating portal for observers and telescopes (S. Crawford)}

The world has numerous public and private astronomical observatories, each with their own suite of instrumentation. The best known publicly operated telescopes, such as Hubble or VLT, typically have significant oversubscription rates, whilst many smaller observatories remain relatively under-used. The idea behind this hack was to create a database of telescopes and instruments, with a web-based front-end for observers to discover what instruments are available for their science; in essence a match-making service for observers and instruments.

\begin{figure}
   \begin{center}
   \begin{tabular}{c}
   \includegraphics[height=8cm]{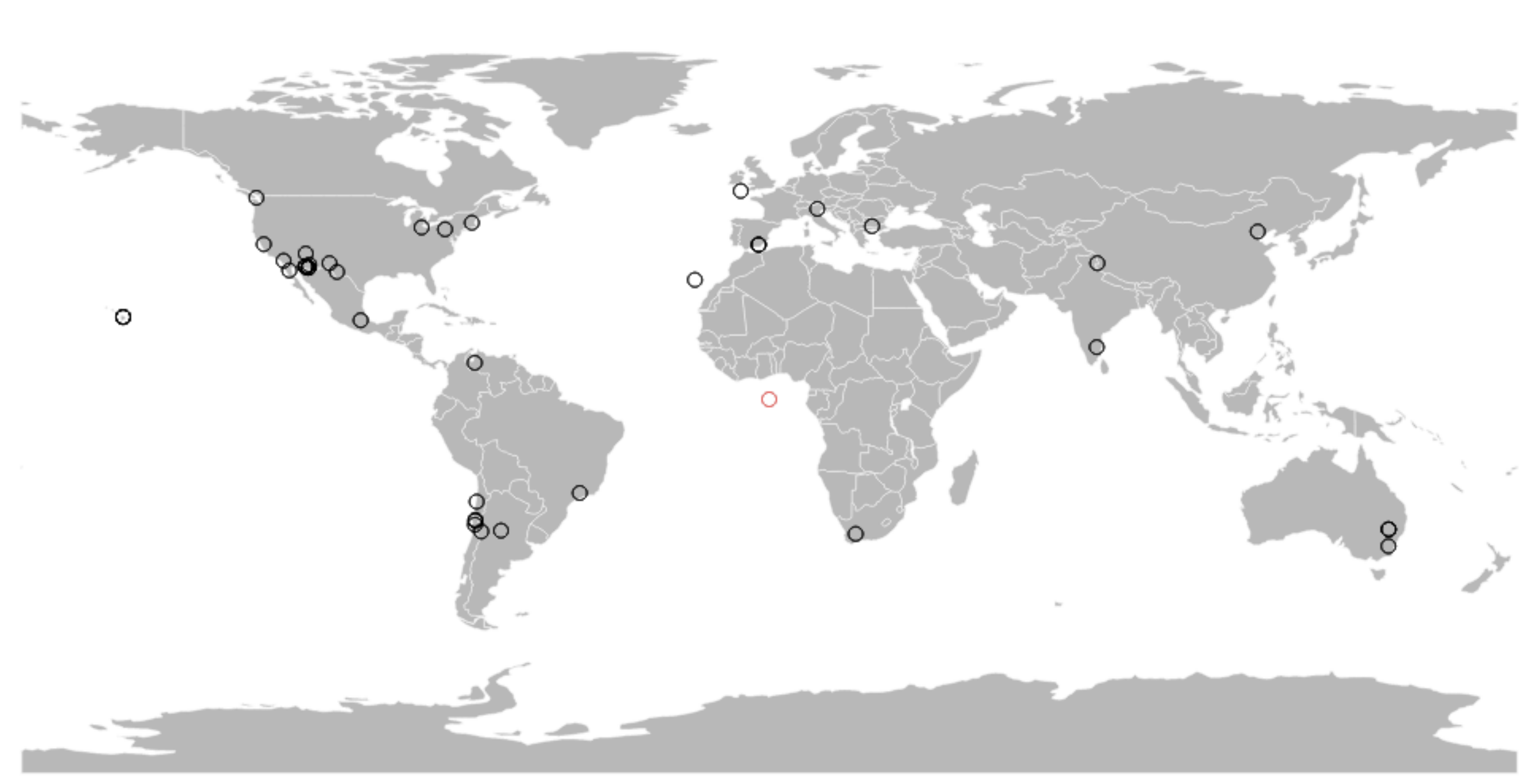}
   \end{tabular}
   \end{center}
   \caption[example]{ \label{fig:sites} Distribution of observing sites around the world created for the OKGalileo hack. The webpage uses d3.js to display observatory site information stored in a Google Spreadsheet. }
\end{figure}

The work for the project was divided amongst the team\footnote{For the full list of contributors, please see \url{https://github.com/jluastro/OKGalileo/graphs/contributors}} into several parts and used several different technologies for displaying the data.  Contributions were coordinated through a github.com repository \footnote{\url{https://github.com/jluastro/OKGalileo}}.  For the database of observatory sites, we used \emph{Google Spreadsheet}.  This  allowed for rapid development of an online form for entering an observing sites information using \emph{Google forms}, development of a Python code for automated entry, and direct access to the database for manual editing.  Sites were automatically uploaded from the \emph{IRAF} database of observatory sites.   With the observatory site data, pages were developed in javascript using the d3.js library for data visualizations.  An example of one of these visualizations is presented in Figure~\ref{fig:sites}, where all of the observatory sites are plotted on a world map.

The one-day hack allowed for the creation of the database, uploading the site information, and creating some basic visualizations of the data. To make the site further useful will require more extensive work, but it does highlight the usefulness of participating in a one day hack. Many of the volunteers on this project were relatively new to using some of the software and were able to become familiar with new resources for building collaborations, data collection, and visualization.

\section{SUMMARY}

We hosted the first SPIE Hack Day at the 2014 Astronomical Telescopes and Instrumentation conference in Montr\'{e}al, with the aim of giving developers and designers to collaborate on innovative projects, perhaps outside of their `regular' professional activities. Approximately 25 conference attendees participated in the event, and we received excellent feedback from a number of these on the organisation and spirit of the Hack Day; the hacks presented covered a wide range of topics, from the development of web-based resources for observational astronomy to more creative sound-based projects.

The Hack Day participants were drawn mainly from authors presenting work in the Software and Cyberinfrastructure conference, however in future the event will be advertised more widely to the entire conference. Experience with previous events has shown that word-of-mouth advertising is an important way of increasing the participant numbers. An exciting component for future SPIE Hack Days will be to enable more hardware-based hacks, perhaps with participation of the exhibitors. Other ideas such as a competitive component with prizes for the best projects may be discussed in this context as well.

\acknowledgements{We would like to thank Gianluca Chiozzi of ESO, as well as Rob Whitner and Amy Nelson at SPIE for their support in organising and publicising the Hack Day event, and all Hack Day participants for their energy, ideas and enthusiasm.

IRAF is distributed by the National Optical Astronomy Observatories, which are operated by the Association of Universities for Research in Astronomy, Inc., under cooperative agreement with the National Science Foundation. Some of the work presented here made use of Astropy, a community-developed core Python package for Astronomy\cite{astropy}}


\bibliographystyle{spiebib}   

\end{document}